\documentclass[fleqn,12pt]{article}
 \usepackage{amsfonts,amssymb}
 \newcommand{\End}{\nonumber\\}

 \newcommand{\Dpath}{{\cal{D}}}
 \newcommand{\half}{\frac12}
 \newcommand{\Intot}{\int_{0}^{t}}
 \newcommand{\Omu}{\omega_{\mu}}
 \newcommand{\Pmu}{p_{\mu}}
 \newcommand{\Xmu}{x^{\mu}}
 \newcommand{\XXmu}{X^{\mu}}
 \newcommand{\Pimu}{\pi_{\mu}}
 \newcommand{\Etamu}{\eta^{\mu}}
 \newcommand{\Tmu}{T_{\mu}}
 \newcommand{\Dmu}{\partial_{\mu}}
 \newcommand{\DDmu}{\nabla_{\mu}}
 \newcommand{\Lag}{{\cal L}}
 \newcommand{\Curv}[4]{R_{#1 #2 #3}{}^{#4}}
 \newcommand{\Emh}{e^{-h}}
 \newcommand{\Eph}{e^{h}}
 \newcommand{\Pb}[2]{\left\{ #1,#2 \right\}}
 \newcommand{\BRST}{{\small BRST}}
 \newcommand{\Gam}[3]{\Gamma_{#1 #2}^{#3}}
\begin{document}
 %
  \begin{flushright}
 KCL-MTH-99-50\\
 December 1999\\
 \
  \end{flushright}
  \begin{center}{\huge Canonical Quantization and Topological Theories}\\
 \ \\
 {\large Talk given at QG99,
 Third Meeting on\\
 Constrained Dynamics and Quantum Gravity,\\
 Villasimius (Sardinia, Italy) September 13-17, 1999\\
 \ \\
 Alice Rogers\\
 Department of Mathematics\\
 King's College}\\
 Strand, London WC2R 2LS, Great Britain\\
  \end{center}
 \begin{abstract}
The canonical quantization of the topological particle is
described; it is shown that \BRST\ quantization of the model gives
the supersymmetric quantum mechanical model considered by Witten
when investigating Morse theory, and the rigorous path integral
method appropriate for this model is discussed. Possibilities for
the extension of this work to two dimensional models are briefly
considered.
 \end{abstract}
%
 \section{Introduction}
In this talk the first steps in a rigorous study of topological
quantum theories using canonical methods is described. A
topological theory is characterised by a high degree of symmetry,
with the number of symmetries equal to the number of fields of the
theory. As a result the system has no dynamical degrees of
freedom, and the space of fields modulo symmetries, which is
trivial at the linear level, is typically a finite-dimensional
moduli space which encodes `topological' information.
 \par
As with quantum theories in general, a powerful  tool for studying
a topological theory is functional integration. The generic
expression for the vacuum generating functional takes the form
 \begin{equation}\label{GFIeq}
  \int_{{\rm fields}/{\rm symmetries}} \Dpath \phi \,
    \exp\left( i S[\phi] \right)
 \end{equation}
where  $S(\phi)$ is the action of the field $\phi$. Using standard
(non-rigorous) methods of quantum field theory a number of new and
unexpected mathematical results have been been derived from
topological models, results which in many cases have then been
fully proved by more standard mathematical methods, but which
would probably not have been discovered without the insights
gained from the quantum field theory. (An early appearance of
topological invariants in the quantum field theoretic situation is
due to Belavin, Polyakov, Schwarz and Tyupin \cite{BelPolSchTyu}.
A more recent example of the powerful application of topological
quantum field theory in mathematics may be found in
\cite{Witten94}, while fuller accounts of earlier work in this
field may be found in the books of Nash \cite{Nash} and Schwarz
\cite{Schwar93}.) Most functional integrals such as (\ref{GFIeq}),
and related expressions with operator insertions, have not at
present been properly defined. However, since these integrals have
such astonishing mathematical power, it seems that an attempt to
define these objects rigorously should be more than worth while.
In this talk we show how this may be done for the simplest
topological model, the topological particle, and describe briefly
some recent work by Hrabak \cite{Hrabak} which might lead to
progress in the canonical quantization of topological field
theories.
 \par
Some rigorous results on path integrals (that is, functional
integrals in quantum mechanics) are known. The basic classical
result (which is described by Simon in \cite{Simon}) for a
particle of unit mass moving in one dimension with Hamiltonian
 \begin{equation}
  H = \half p^2 + V(x)
 \end{equation}
gives the action of the imaginary time  evolution
operator $\exp(-Ht)$ on a wave function $\psi(x)$ by the formula
 \begin{equation}\label{EVeq}
   \exp (-Ht) \psi(x)
  = \int d\mu \exp \left( -\int_0^t V((x(s)) ds \right) \psi(x(t))
 \end{equation}
where $d\mu$ denotes Wiener measure starting from $x$, and $x(t)$
are corresponding Brownian paths; the potential $V$ must satisfy
certain analytic conditions. The curved space analogue of this
result for a Riemannian manifold has been developed by Elworthy
\cite{Elwort} and by Ikeda and Watanabe \cite{IkeWat}. The
expression for evolution according to the Hamiltonian $H= L +V(x)$
where $L$ is the scalar Laplacian looks identical to (\ref{EVeq}),
but with $x(t)$ a process depending on metric and connection
rather than simply flat space Brownian motion. Tangent space
geometry plays an essential part in the theory. The present author
has further extended these methods  by developing a flat space
theory of fermionic path integrals \cite{GBM} and marrying it with
Brownian motion on manifolds to give Brownian motion on
supermanifolds in a suitable form for handling  the Hodge-de Rham
operator and the Dirac operator on manifolds \cite{SCSTWO,JMPFI}.
 \section{The Topological particle}
Following Beaulieu and Singer \cite{BeaSin} we consider the
quantum-mechanical model with fields which are maps $x: I \to M$,
where $I$ is the interval $[0,t]$ and $M$ is an $n$-dimensional
Riemannian manifold with metric $g$. The action of the theory is
 \begin{equation}\label{ACeq}
  S[x(.)] = \Intot i\Omu(x(t'))\dot{x}^{\mu}(t') \, dt'
 \end{equation}
$\omega=dh$ is an exact one form on $M$ with local co-ordinate
expression $\omega= \Omu(x) d\Xmu$ and
 $\dot{x}^{\mu}(t')= \frac{d{x}^{\mu}}{d t'}$. The action can
be expressed in the simpler form $S[x(.)] = i(h(x(t))-h(x(0)))$ which
shows that the action is indeed highly symmetric, being
independent of all but the endpoints of the paths $x(t)$.
 \par
While Beaulieu and Singer consider the case where $\omega=0$, we
consider the case where $h$ is a Morse function on $M$ (so that
$\omega=dh$ is only zero at isolated points). To carry out the
canonical quantization we first evaluate the momentum $\Pmu$
conjugate to $\Xmu$, obtaining
 \begin{equation}\label{MOMeq}
  \Pmu = \frac{\delta \Lag}{\delta \dot{x}^\mu} = i\Omu,
 \end{equation}
which shows that the theory has $n$ constraints
 \begin{equation}
   \Tmu \equiv \Pmu - i\Omu.
 \end{equation}
The Hamiltonian of the theory is as usual defined to be
 $H(p,x)= \Pmu \Xmu - \Lag(x,\dot{x})$, so that, as is generally
the case for a topological theory, the Hamiltonian of the theory
(prior to gauge-fixing) is zero. Since $\omega$ is closed these
constraints are first class and abelian, that is
$\Pb{\Tmu}{T_{\nu}}=0$ and $\Pb{\Tmu}{H_c}=0$.
 \par
The infinitesimal gauge transformations generated by $\Tmu$ are
 \begin{equation}\label{GTeq}
  \delta_{\epsilon}\psi(x) =-i \epsilon(\Dmu \psi(x) + \Omu(x) \psi(x))
 \end{equation}
where $\psi(x)$ is a wave function and  quantization is in the
Schr\"odinger picture with $\Pmu =-i \Dmu$.  (Below, when ghosts are
introduced, we will find that we require $\Pmu$ to be represented
as a covariant derivative $-i\DDmu$.) The explicit form of the gauge
transformations  suggests that
representative of gauge equivalence classes may be obtained from
the condition $\XXmu \psi=0$ where
 $\XXmu=g^{\mu\nu} (p_{\nu}  + i\omega_{\nu})$. The validity of these
gauge-fixing conditions will become clear below.
 \par
The \BRST\ quantization scheme will now be applied to this system;
to do this anticommuting ghosts  $\Etamu$ together with their
conjugate momenta $\Pimu$ are introduced. (The phase space is now
a $(2n,2n)$-dimensional supermanifold, with odd coordinates
$\Etamu,\Pimu$ transforming as indices suggest.) Poisson brackets
on this extended phase space are defined by the symplectic form
 \begin{equation}\label{SPBeq}
  d\Pmu \wedge d \Xmu + \nabla\Pimu \wedge \nabla\Etamu
   +  \frac12 dx^{\mu} \wedge dx^{\nu}
  \Curv{\mu}{\nu}{\kappa}{\lambda}\eta^{\kappa}\pi_{\lambda},
 \end{equation}
(where $\nabla$ denotes covariant differentiation using the
Levi-Civita connection); this is a special case of the symplectic
form introduced by Rothstein \cite{Rothst91}. Quantization
 is carried out by introducing states represented
by wave functions $\psi(x,\eta)$ and momenta acting as
 $\Pmu =-i \DDmu$, $\Pimu = -i\frac{\partial}{\partial\Etamu}$.
The wave functions $\psi(x,\eta)$ are functions on the
$(n,n)$-dimensional supermanifold $SM$ with local coordinates
$\Xmu,\Etamu$, and the explicit form of the action of the
covariant derivative on a wave function is given by
 \begin{eqnarray}\label{CDeq}
  \DDmu \psi(x,\eta) = \Dmu \psi(x,\eta)
     + \Gam{\mu}{\nu}{\lambda} \eta^{\nu}
     \frac{\partial}{\partial\eta^{\lambda}}\psi(x,\eta).
 \end{eqnarray}
 \par
The \BRST\ charge $Q$ takes the standard form
 $Q=\Etamu\Tmu=-i\Etamu(\Dmu + \omega)$. (A covariant derivative is
not required here because of  the symmetry of the connection.) The
gauge-fixing fermion $\chi$ also takes the standard form
 $\chi = \Pimu\XXmu = -ig^{\mu\nu} \Pimu(\nabla_{\nu}-\omega_{\nu})$.
Alternatively, if we make the natural identification of forms on
$M$ with wave functions $\psi(x,\eta)$ then  $Q=-i\Emh d \Eph$,
$\chi = \Eph \delta \Emh$ where $d$ denotes exterior
differentiation of forms and $\delta= *d*$ is the adjoint
operator. The gauge-fixing Hamiltonian is then
 \begin{eqnarray}\label{HAMeq}
 H_g &=& i( Q \chi + \chi Q) \End
 &=& d \delta + \delta d + g^{\mu\nu}\Omu \omega_{\nu}
  -i (\Pimu\eta^{\nu} -
 \eta^{\nu}\Pimu) \frac{\partial^2 h }{\partial \Xmu \partial x_{\nu}}.
\end{eqnarray}
 \par
This Hamiltonian has appeared in the literature on  other
occasions; for instance, when $h$ is constant, it is the
Hamiltonian used by Alvarez-Gaum\'e \cite{Alvare} to prove the
Atiyah-Singer index theorem. (A rigorous version of this proof may
be found in \cite{SCSTWO}.) It is also the first supersymmetric
Hamiltonian used by Witten in his study of Morse theory
\cite{Witten82}.
 \par
It is  evident that the choice $\chi=\Pimu\XXmu$ is a good
gauge-fixing condition, satisfying the  essential conditions
derived by the author in \cite{GFBFVQ}. First, as observed by
Beaulieu and Singer in the constant $h$ case, the standard theory
of harmonic forms shows that the gauge condition determines a
unique element of each $Q$ cohomology class. (The observation that
these arguments extend to all functions $h$ is due to Witten
\cite{Witten82}.) Also, the zeros of $H_g$ coincide with these
representatives of the cohomology classes, while the eigenvalues
of $H_g$ tend to infinity, so that this Hamiltonian does regulate
the non-physical states.
 \par
The path integral formulae for this Hamiltonian can be put in
rigorous form using the methods of the author in
\cite{SCSTWO,JMPFI}, and used to establish rigorous results for
this model. The key idea in this approach is to use Brownian paths
$x_t,\eta_t$ in the supermanifold $SM$ with local coordinates
$\Xmu,\Etamu$. The Brownian paths are defined by the stochastic
differential  equations
 \begin{eqnarray}
  \Xmu_t &=& \Xmu + \Intot e^{\mu}_{a,s}\circ db^{a}_s , \End
  e^{\mu}_{a,t}&=& e^{\mu}_{a} +\Intot
   -e^{\nu}_{a,s} e^{\lambda}_{b,s}
   \Gam{\nu}{\lambda}{\mu}(x_s) \circ db^{b}_s \End
  \Etamu_t &=& \Etamu + \theta^{a}_t e^{\mu}_{a,t} \End
  +&&\!\!\!\!\!\!\!\!\! \Intot \big( - \eta^{\nu}_s \Gam{\nu}{\lambda}{\mu}
  e^{\lambda}_{b,s} \circ db^{b}_s - \theta^a_t de^{\mu}_{a,s}
  +\frac{1}{4}\eta^{\nu}_s \Curv{\nu}{\lambda}{\kappa}{\mu}
  (x_s)\eta^{\lambda}_s\rho^{a}_s e^{\kappa}_{a,s} ds \big),
 \end{eqnarray}
where $b_t$ is flat bosonic Brownian motion and $(\theta_t,\rho_t)$
is flat fermionic Brownian motion. The measure corresponding
to this process incorporates as the `kinetic term' the heat kernel
of the Laplace-Beltrami operator $(d+ \delta)^2$, so that these
Brownian paths are appropriate for the analysis of the Hamiltonian
(\ref{HAMeq}). Further details will be found in \cite{TOPP}, where
it will be shown that Witten's approach to Morse theory
\cite{Witten82} can be put on an entirely rigorous mathematical
footing. (Some parts of Witten's analysis have been proved
rigorously by Simon et al \cite{Simetal} and by Mathai and Wu
\cite{MatWu}; however the explicit modeling of the manifold's
cohomology via critical points and instanton calculations does not
appear to have received a full mathematical treatment.)
 \section{The two-dimensional topological sigma \hfil\break model}
To conclude, a brief indication of some developments in a
two-dimensional topological model will be described. The model,
which was first proposed by Witten \cite{Witten88}, concerns the
geometry of $J$-holomorphic curves (or pseudoholomorphic maps)
$u:\Sigma \to M$  from a Riemann surface $\Sigma$ into an a
$2m$-dimensional almost-Kaehler manifold $M$. The other fields of
the theory are a bosonic set $H^{\alpha}_{\mu}$ and two fermionic
sets $\Etamu$ and $\pi^{\alpha}_{\mu}$. (Here $\alpha=1,2$ are
indices on $\Sigma$ while $\mu=1,\dots,2 m$ are indices on $M$.)
The fields $H$ and $\pi$ satisfy constraints
 $P^{^{-}}H=0, P^{^{-}}\pi=0$ where
 $P^{^{-}}{}^{\alpha\mu}_{\beta\nu}
 =\delta^{\alpha}_{\beta}\delta^{\mu}_{\nu}-
  \epsilon^{\alpha}_{\beta}J^{\mu}_{\nu}$
is a projection operator with $\epsilon$ the complex structure on
$\Sigma$ and $J$ the almost complex structures on $M$. Witten
constructs by hand a set of supersymmetry transformations
(beginning with $\delta u^{\mu} = i \epsilon \eta^{\mu}$) on these
fields, and then an invariant action. The model is used to derive
deep geometric insights into the moduli space of $J$-holomorphic
curves on $M$. Recent work of Hrabak \cite{Hrabak} shows that the
rather complicated and seemingly ad hoc supersymmetry
transformations of the model can be derived in the canonical
setting as \BRST\ transformations; the novel feature of Hrabak's
work is that the formalism used is not the standard canonical
formalism (in which time plays a special r\^ole) but the
multisymplectic formalism which is manifestly covariant;
corresponding to the fields $u:\Sigma \to M$ there are
multimomenta $p^{\alpha}_{\mu}$ (which after projection relate to
Witten's  $H^{\alpha}_{\mu}$), while in the ghost sector the
ghosts have momenta ${\cal{P}}_{\mu}^{\alpha}$ which relate to
Witten's $\pi^{\alpha}_{\mu}$. The BRST symmetry obtained by
Hrabak corresponds directly to the $J$-holomorphicity of the
embeddings. Recent work by Kanatchikov \cite{Kanatc} on
quantization in the multisymplectic framework suggests that it may
be possible to use Hrabak's approach to carry out a full canonical
quantization of Witten's interesting two-dimensional topological
model.

\end{document}